\documentclass[final,5p,times]{elsarticle}

\journal{Scripta Materialia}

\usepackage{graphicx}%
\usepackage{dcolumn}%
\usepackage{bm}%
\usepackage{miller} %
\usepackage{lineno} %
\usepackage[]{textgreek}
\usepackage{siunitx}

\newcommand{\MAX}{Ti$_3$AlC$_2$}
\newcommand{\TiC}{Ti$_\text{x}$C$_\text{y}$}
\newcommand{\ie}{\emph{i.e.}}
\newcommand{\etal}{\emph{~et al.}}

\begin{document}

\begin{frontmatter}

\title{Features of a nano-twist phase in the nanolayered \MAX\ MAX phase}

\author[lem3,labex]{Julien Guénolé}
\author[lem3,labex]{Vincent Taupin}

\affiliation[lem3]{organization={Université de Lorraine, CNRS, Arts et Métiers, LEM3},
            city={Metz},
            postcode={57070},
            country={France}}

\affiliation[labex]{organization={Labex DAMAS, Université de Lorraine},
            city={Metz},
            postcode={57070},
            country={France}}

\author[lms,ls]{Maxime Vallet}
\affiliation[lms]{organization={Laboratoire Mécanique des Sols, Structures et Matériaux, CentraleSupélec, CNRS UMR 8579, Université Paris-Saclay},
            city={Gif-sur-Yvette},
            postcode={91190},
            country={France}}
\affiliation[ls]{organization={Laboratoire Structures, Propriétés et Modélisation des Solides, CentraleSupélec, CNRS UMR 8580, Université Paris-Saclay},
            city={Gif-sur-Yvette},
            postcode={91190},
            country={France}}         

\author[cmse]{Wenbo Yu}
\affiliation[cmse]{organization={Center of Materials Science and Engineering, School of Mechanical and Electronic Control Engineering, Beijing Jiaotong University},
            city={Beijing},
            postcode={100044},
            country={China}}  

\author[lem3,labex]{Antoine Guitton\corref{cor1}}
 \cortext[cor1]{antoine.guitton@univ-lorraine.fr}

\begin{abstract}
Complex intermetallic materials known as MAX phases exhibit exceptional properties from both metals and ceramics, largely thanks to their nanolayered structure.
With high-resolution scanning transmission electron microscopy supported by atomistic modelling, we reveal atomic features of a nano-twist phase in the nanolayered \MAX.
The rotated hexagonal single-crystal is encompassed within basal symmetric twist interfaces similar to grain boundaries. In particular, we show that air-oxidation at \SI{1000}{\celsius} can form a twisted phase that leads to the formation of interfacial dislocation networks with screw characters or to severe interfacial reconstructions. Additionally, we explore the contribution of disclinations to the representation by continuum models of the stress field generated by such nano-twist defect in the \MAX{} bulk phase. The occurrence of this unexpected defect is expected to impact the physical response of this nanolayered-based material as such supports property-by-design approaches.
\end{abstract}

\end{frontmatter}

\MAX{} belongs to the wide family of MAX phases~\cite{Wang2002SolidliquidCeramic}. These ternary compounds arise a keen interest, because they are stiff, lightweight, machinable, made from relatively inexpensive raw materials, resistant to oxidation and thermal shock, and capable of remaining strong up to temperatures above 1,300°C in air~\cite{Barsoum2001TheMaterials}. More than fifty compounds are thermodynamically stable and all of them exhibit the same range of promising properties. Because of their composition, they were called M$_{n+1}$AX$_n$ phases ($n =$ 1 to 3, M is a transition metal, A is an A-group element and X is nitrogen and/or carbon)~\cite{Barsoum2013MAXPhases}. 

As typical MAX phase, \MAX{ } has a nanolayered structure with an hexagonal lattice, whose the space group is $P6_3/mmc$~\cite{Barsoum2001TheMaterials,Barsoum2000TheSolids}. The primitive cell can be described as a stacking of two Ti$_6$C octahedron layers with one layer of Al~\cite{Wang2002SolidliquidCeramic}. Furthermore, measurements of lattice parameters with numerous methods reveal that \MAX{} exhibits elevated high crystalline anisotropy. The $c/a$ ratio is slightly higher than 6~\cite{Barsoum2011ElasticPhases}. 

It is established, that MAX phases experience plastic deformation by the glide on basal planes of a-dislocations confined between the M$_6$X and Al layers, thus forming pile-ups and walls~\cite{Farber2005DislocationsTi3SiC2,Barsoum2011ElasticPhases,Gouriet2015DislocationModel}. The latter can form local disorientation areas, known as kink bands~\cite{Barsoum1999DislocationsTi3SiC2,Guitton2014EffectDiffraction}. Furthermore, numerous dislocation interactions forming networks (dipoles, reactions) and high lattice friction have been observed~\cite{Guitton2012DislocationPressure, Bei2013Pressure-enforcedInvestigation}. Note, that out-of-basal plane dislocations have been observed as well, but they do not play a role at room temperature in standard deformation conditions \cite{Tromas2010c,Joulain2008}. At high temperature, it has been revealed, that out-of-basal plane \hkl<a>-dislocations are common events and hence cross-slip (from basal planes to prismatic or pyramidal planes) plays a key role in the deformation~\cite{Guitton2015EvidenceTemperature,Drouelle2017DeformationPhase}. This increase of available glide systems is likely to originate the brittle-to-ductile transition of MAX phases~\cite{Guitton2015EvidenceTemperature}. Note also that Frank partial \hkl<c>-dislocations correlated with a diffusion mechanism were recently reported \cite{Yu2021FrankDiffusion}. Moreover, observations of stacking faults is another major microstructural feature, but their role in deformation both at room and high temperatures remains unclear \cite{Farber2005DislocationsTi3SiC2,Joulain2008,Drouelle2017DeformationPhase}.

Of all MAX phases, those containing Al, such as \MAX, are the most resistant to oxidation \cite{Barsoum2001TheMaterials}. During exposure of Ti--Al--C MAX phases to oxidizing environment at high temperatures, the outward diffusion of the weakly bonded Al atoms is much faster compared to the more covalently bonded Ti atoms~\cite{Barsoum2001TheMaterials,Drouelle2020OxidationComparison,Drouelle2020Microstructure-oxidationPhase,Yu2020OxidationSize}. Therefore, it results in a regime, where a superficial protective layer of Al$_2$O$_3$ is formed. However, TiO$_2$ can be formed as well, leading, this time, to a catastrophic regime. Both mechanisms are strongly depend on the oxidation conditions and the initial microstructure~\cite{Drouelle2020Microstructure-oxidationPhase,Yu2020OxidationSize}. Despite the description of these oxides, operating oxidation mechanisms of \MAX{} and more generally of MAX phases remain poorly documented, especially on how the crystallographic structure is evolving.
During decomposition of 312 MAX phases (such as Ti$_3$AlC$_2$, Ti$_3$SiC$_2$) at high temperatures in an oxygen-containing atmosphere or in species having a high affinity to A-element, A-element is observed to de-intercalate, thus forming Ti$_3$C$_2$ platelets \cite{Zhang2007StructureComposites, Wang2003StabilityArgon, Emmerlich2007ThermalFilms, Barsoum1999TheCryolite}. More precisely, in case of Ti$_2$AlC-Cu composite, this de-intercalation of Al is associated with a Frank partial dislocation-based mechanism~\cite{Yu2021FrankDiffusion}. This formation of \TiC{ }platelets is then viewed as reinforcement of the composites~\cite{Zhang2007StructureComposites}.

Such is the background of the current letter. This work is motivated by the analysis of original phases that were observed after oxidation in a \MAX\ matrix by high-resolution scanning transmission electron microscopy (HR-STEM). Such nanolamellar phases appear to be twisted with respect to the surrounding matrix and also associated with diffusion mechanisms in its neighborhood. Hence, by interpreting our HR-STEM observations using molecular dynamics simulations of model nano-twist phases, we study their atomic scale features including interfacial crystal defects, excess energy, and mechanical fields. Concerning the latter, a continuous description of the twist phase using a disclination based mechanical framework is conducted complementary to atomistic simulations to further evidence the complexity of such phases.

The specimen was prepared by hot isotropic pressing (HIP). Briefly, powders of Ti, Al and TiC were mixed in stoechiometric proportions 2Ti:1.05Al:0.85C and then cold-compacted into cylindrical steel dies using an uniaxial pressure. Powder compacts were encapsulated into pyrex containers under high vacuum for reactive sintering in the HIP machine. Afterwards, the specimen were oxidized at \SI{1000}{\celsius} for \SI{25}{\hour} in a flowing air atmosphere. More details are given in~\cite{Yu2014,Yu2020OxidationSize}.
Cross-sectional TEM samples were prepared by focused ion beam (FIB) on a FEI Helios dual beam Nanolab 600i. Atomically-resolved high-angle annular dark field scanning TEM (HAADF-STEM) was performed on an Titan$^3$ G2 S/TEM fitted with a double aberration corrector for probe and image corrections and operating at \SI{300}{\kilo\volt}.

HR-STEM in HAADF mode of the lamella is shown in Fig.~\ref{fig:Exp} and at lower magnification in Fig.~SM1 in supplementary materials. As HAADF detector senses a greater signal from atoms with a higher atomic number Z, Ti columns appear brighter in the resulting micrograph~\cite{Williams2009TransmissionMicroscopy}. The nano-layered structure of \MAX{} highlighted in the circular inset is consistent with the expected one for a 312-MAX phase projected along \hkl[-1 -1 2 0] \ie{} zigzag stacking of three TiC planes followed by one Al plane, the Al plane being the axis of symmetry of the zigzag. 
Several large \TiC{} laths are clearly visible as well. These sub-phases are probably similar to Ti$_3$C$_2$ phases as they are formed by the diffusion of the Al interlayers of the \MAX{} phase. However, the experiments we used are not able to characterize them precisely, and the mechanisms responsible for their formation is out of the scope of this letter. The region of interest shows blurred atomic layers (ROI, white square in Fig.~\ref{fig:Exp}). The blur is along the \hkl[1 -1 0 0] direction, localized within the basal plane. This indicates that this part of the lattice has undergone a rotation around the \hkl[0 0 0 1] direction. The boundaries between this rotated phase and the other phase are marked with dashed white lines. 
Particular orientation relationships for MX platelets in MAX interfaces have been reported in the literature~\cite{Ma04PM, Lin07JMST}. However, such orientations will not produce the high-resolution contrast we observe in Fig.1, in particular, the blurred atomic layer within the NTP.
Note, that the upper extremity of the ROI shows a large, blurred area and visible inter-diffusion between Al and Ti planes. 
Such mechanism that induce severe local lattice strain might be responsible for the difference in the contrast of the \TiC{} phase on both side of the NTP (Fig.~\ref{fig:Exp}).

\begin{figure}[t]
\centering
\includegraphics[width=0.99\columnwidth]{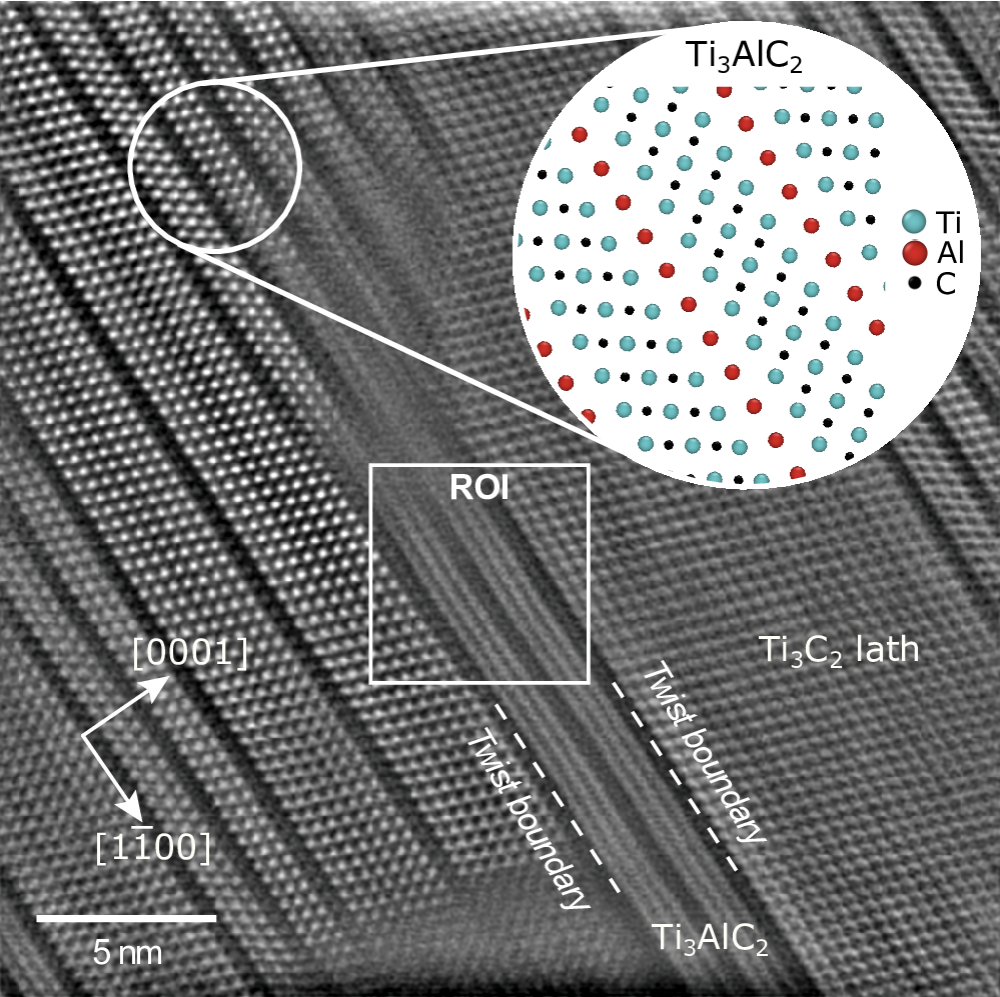}
\caption{\label{fig:Exp} Experimental observation of the nano-twist phase (region of interest, ROI) in the nanolayered \MAX. Filtered HR-STEM micrography in HAADF mode with electron direction along \hkl[1 1 -2 0]. Boundaries of the nano-twist phase are indicated by dashed lines. A model of the \MAX\ crystallographic structure is shown in the inset. Some \TiC{} similar to Ti$_3$C$_2$ laths are also observed.}\end{figure}

\begin{figure}[t]
\centering
\includegraphics[width=0.99\columnwidth]{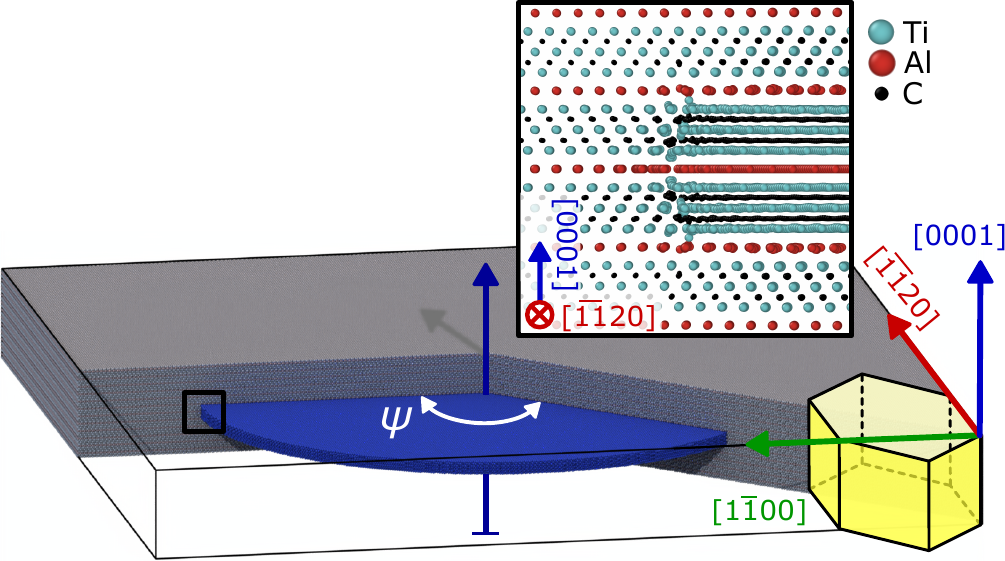}
\caption{\label{fig:SimSample} Modeling of the nano-twist phase in the nano-layered \MAX\ by means of a cookie-shape rotation region. 
(a) atomistic simulation setup of the cookie-shape region (in blue) parallel to the basal planes, with \hkl[0001] rotation axis and $\psi$ rotation angle. 
The black square indicate the location of the inset.
: magnified projected view of a \SI{10}{\nano\meter} thin slice of the side of the cookie-shape nano-twist phase for $\Psi = \ang{3}$. Ti, Al and C are colored in green, red and black, respectively.
}
\end{figure}

To gain more insights on the observed defect and to characterize precisely the structure of the twist boundary, we modeled the nano-twist phase by atomistic simulations. 
Molecular dynamics (MD) based on interatomic potentials is an excellent modelling method to investigate the atomic-scale configuration of defects, in particular interfaces~\cite{Priester13, Weinberger2016}. It has been widely used over the past decade to investigate grain boundaries~\cite{Sansoz2005, Frolov2013EffectBoundaries, Tschopp2015, Zhao2020InterplayAlloy} and phase boundaries~\cite{Mishin04AM, Prakash15AM, Vaid19AM, Guenole21MD}. 
The only published model suitable to describe \MAX{} atomic interactions is the bond-order potential (BOP) recently adjusted by Plummer and Tucker~\cite{Plummer2019Bond-orderPhases}. It is based on the formalism initially proposed by Tersoff~\cite{Tersoff88PRB} and described in detail by Albe\etal~\cite{Albe02PRB}.
Fig.~\ref{fig:SimSample} shows the atomistic simulation setup to compute the energy and the structure of the rotated phase boundaries. Within a fully periodic bulk \MAX\ phase, a cookie shape region in the center is rotated with an angle $\Psi$ around an axis normal to the basal plane, as indicated in blue in Fig.~\ref{fig:SimSample}. The inset shows as an example a rotation of the atomic structure for $\Psi = \ang{3}$. Note that structures with similar contrasts to HAADF-STEM ones can be obtained with different values of $\Psi$ (Fig.~\ref{fig:Exp}).

It is important to mention that we do not intend here to model the exact system observed in our experiments, for several reasons. (1) The limitation of our experimental approaches do not reveal the exact composition of the \TiC{} phase. (2) The interaction potential used in our MD simulations has been designed to model the \MAX{} phase and, thus, might be less reliable for any non-stochiometric \TiC{} phase. (3) The local atomic environment of the interface, up to three TiC interlayers, is identical for both \MAX{} and \TiC{} phases.
In this context, changing the material that forms the interfaces of the NTP in our MD simulations should influence the value of the interfacial energy, but will have a negligible impact on what is the focus of our work: the energetic profile and the crystallographic structure of the interfaces.
The consideration of idealized systems or surrogate materials is common practice with atomistic simulations, including for direct comparisons with experimental results~\cite{Prakash15AM, Chang2019,Guenole21MD,Yu2021FrankDiffusion}. 

\begin{figure}[t]
\centering
\includegraphics[width=0.99\columnwidth]{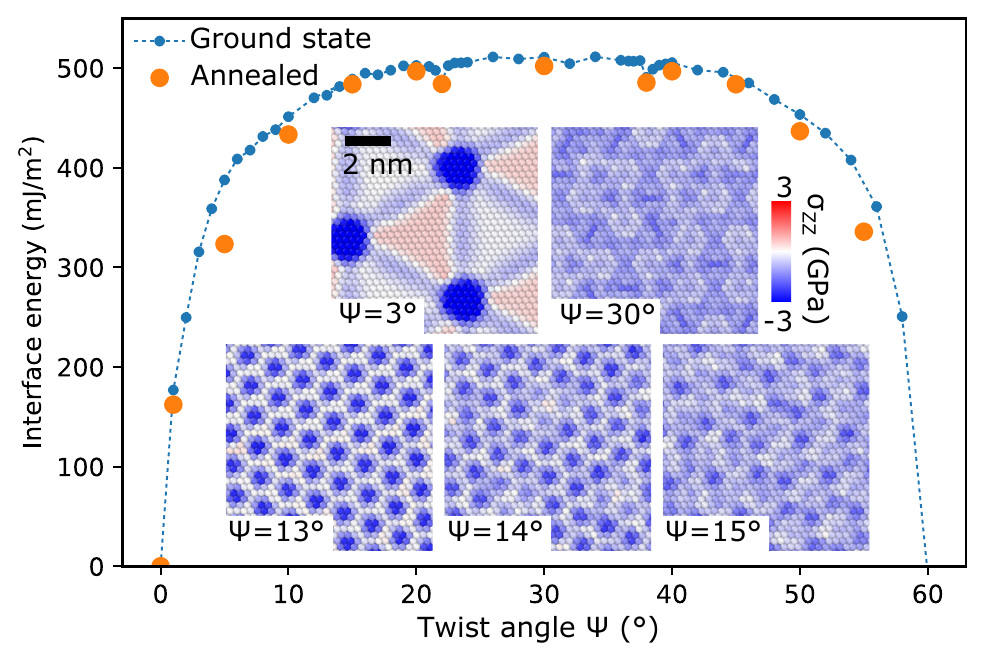}
\caption{\label{fig:InterfaceEnergy} Energy of the basal interface of the nano-twist phase as function of the twist angle $\Psi$. Interfacial energy of the ground state and annealed structure in blue and orange dots, respectively. The dotted line is a guide for the eyes. Insets show magnified views of the stress field $\sigma_{zz}$ for $\Psi = 3, 13, 14, 15, 30$.}
\end{figure}

The as-formed nano-twist phase (NTP) exhibits prismatic and basal interfaces with the \MAX{} phase. The prismatic interface is not clearly defined in our experimental observation, whereas the basal interface appears atomically sharp (See Fig.~\ref{fig:Exp}). In this work, we thus focus on the characterisation of the NTP basal boundary (NTB). 
Fig.~\ref{fig:InterfaceEnergy} shows the interfacial energy of the NTB as function of the twist angle $\Psi$. The energy is computed by considering a cylindrical region in the center of the NTP, that does not encompass the prismatic boundaries. For each twist angle, the system is statically relaxed by using conjugate gradients and fire~\cite{Bitzek2006,Guenole2020AssessmentLAMMPS} methods. The dimensions of the box are adapted to ensure a globally stress-free system.
The energy of such as-relaxed NTB is shown with blue small dots in Fig.~\ref{fig:InterfaceEnergy}.
Selected configurations have been annealed at \SI{600}{\kelvin} for \SI{100}{\pico \second}. A Nose-Hoover thermostat control the temperature and a Nose-Hoover barostat ensure constant zero pressure. The systems are subsequently quenched and their energies are shown in Fig.~\ref{fig:InterfaceEnergy} with orange large dots.

\begin{figure}[t]
\centering
\includegraphics[width=0.99\columnwidth]{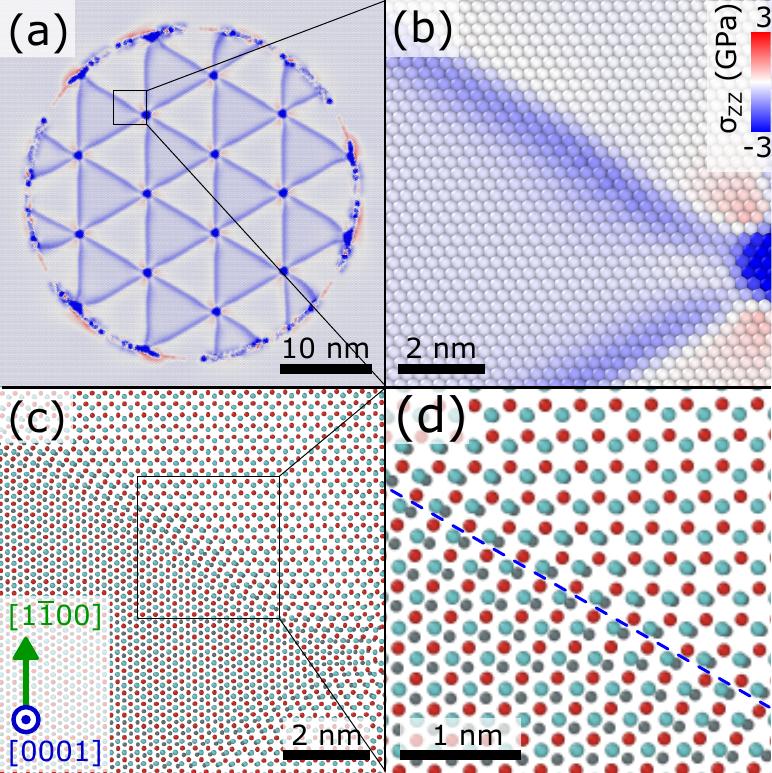}
\caption{\label{fig:LowAngleTB} Interfacial structure of the \MAX\ nano-twist phase for $\Psi = \ang{1}$. (a) global and (b) magnified views of the local stress field $\sigma_{zz}$ revealing the typical dislocation network of a low angle grain boundary. (c) identical view as (b) but showing chemical species as in Fig.~\ref{fig:SimSample}. (d) magnified view of (c) revealing the atomic displacements induced by a screw dislocation. (c) Atomic radius are proportional to the stacking position in the \hkl[0001] direction: light green, red and dark green, from top to bottom. The screw dislocation line is indicated as a blue dashed line.}
\end{figure}

The insets in Fig.~\ref{fig:InterfaceEnergy} present the out-of-plane stress component $\sigma_{zz}$ of the as-relaxed NTB for different $\Psi$. By showing the distortion of the crystal lattice at the basal boundary, it reveals the structures of the interface.
More details are presented in Fig.~\ref{fig:LowAngleTB} for $\Psi = \ang{1}$, the smallest twist angle considered in this work. 
The distortion of the crystal lattice at the NTB is captured by the out-of-plane stress component $\sigma_{zz}$ (Fig.~\ref{fig:LowAngleTB}a,b). This evokes the signature of an interfacial dislocation network, with the three-fold symmetry of the basal plane and the nodes of interacting interfacial dislocations. 
The atomic positions for different basal interlayers shown in Fig.~\ref{fig:LowAngleTB}(c)(d) confirm this observation, as the relative displacement of the atoms is characteristic of screw dislocations.

The NTP we observed experimentally is not a grain on itself, but is located within one crystallographically homogeneous area containing perfectly coherent \MAX{} and \TiC{} phases. The phase \TiC{} being directly issued from the \MAX{} phase, this area can be considered as a \MAX-base grain.
This nano-phase exhibits limited degree of freedom, with its twist axis and at least one interface being precisely defined: the \hkl<c> axis and the basal planes, respectively.
It is clear that the basal interface, we called NTB, shares similarities with pure twist grain boundaries, such as 
(1)~an interfacial energy evolution with a typical bell-shape, 
(2)~an interfacial structure that can be described by a network of screw dislocations lying on basal planes and cross-sliping to 1st-order prismatic planes (consistent with previous experimental observations \cite{Guitton2015EvidenceTemperature}) for low twist angles, 
(3)~a spacing between interfacial dislocations inversely proportional to low twist angle values and 
(4)~a description of the interface in term of dislocations network that vanished for high twist angles. The transition from a low angle configuration to a high angle configuration is for $\Psi \approx \ang{15}$ (Fig.~\ref{fig:InterfaceEnergy}, insets).
However, the NTB also exhibits clear differences towards grain boundaries, namely 
(1)~only one macroscopic degree of freedom (DoF) instead of five for grain boundaries and 
(2)~interfacial reconstructions for high twist angle leading to non-monotonous structural patterns. 

Interestingly, such NTP were already reported by Drouelle \emph{et al.} in \MAX{} deformed by creep at high temperature. Indeed, they observed by conventional TEM, highly disorganized lenticular defects with a very high density of screw dislocations \cite{Drouelle2017DeformationPhase}. In addition, Zhang \emph{et al.} reported low-angle twist grain boundaries (disorientation around \SI{0.5}{\degree}) in \MAX{} compressed uniaxially at \SI{1200}{\celsius} \cite{Zhang2016}. Such sub-grain boundaries are formed by screw dislocation networks originating from basal dislocation reactions, like those predicted here in Fig.~\ref{fig:LowAngleTB}.a. Both studies conclude that such features may play a role in the plasticity of \MAX.

It is worth noticing that, within the time frame considered by our atomistic simulations, neither the interfacial energy nor the interfacial structure is significantly altered by annealing. The configuration of the NTB predicted by atomistic modelling appears as such stable. 

\begin{figure*}[t]
\centering
\includegraphics[width=0.99\textwidth]{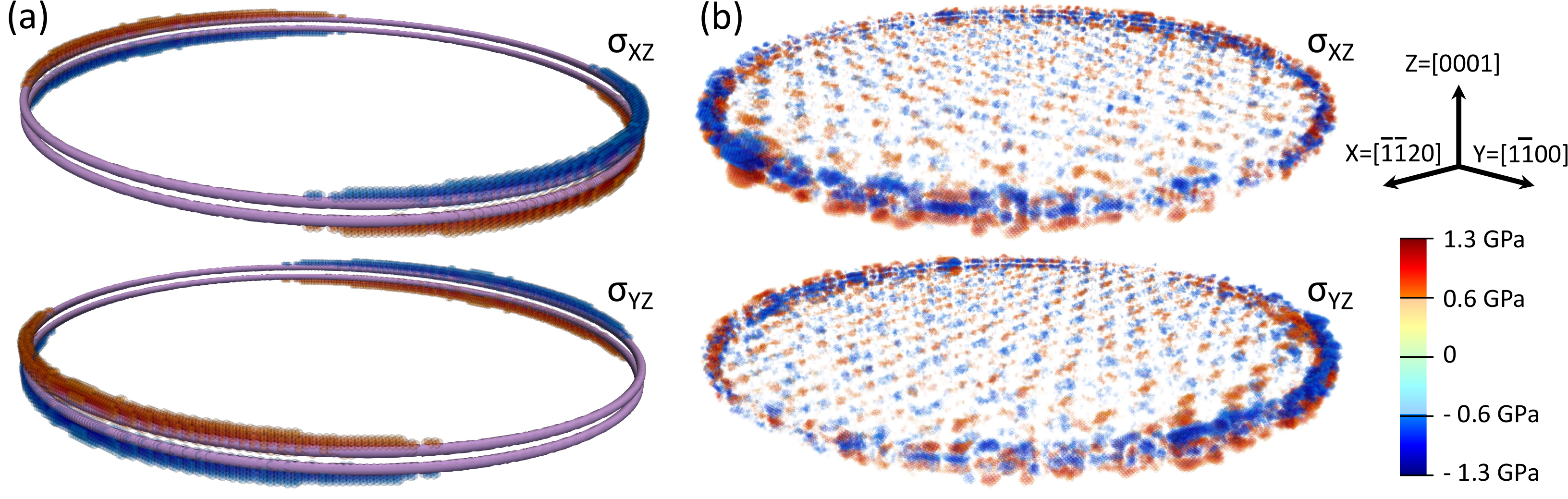}
\caption{\label{fig:DisclinLoops} Shear stress fields $\sigma_{xz}$ and $\sigma_{yz}$ of a high-angle nano-twist phase represented by (a) a dipole of twist disclination loops and (b) an atomistic cookie-shape region. Only  voxels (a) and atoms (b) with stresses with magnitude larger than 600 MPa are shown. In purple (a) is shown the norm of the disclination density tensor.} 
\end{figure*}

For high twist angles ($\Psi > \ang{15}$), the interfacial configuration exhibits severe reconstructions. By comparison to configurations with low twist angles, this reconstruction releases localized atomic stresses but does lead to lower interfacial energies. 
Interfacial energies for tilt and twist boundaries in crystals with cubic structures often show clear drops for particular $\Psi$ values that correspond to high symmetry misorientations. Yet, such stable high angle configuration are not observed with the NTB investigated in this work. This can be related to the limited DoF accessible to this boundary, as described in the following. 
Grain boundaries are classically considered within the coincidence site lattice theory (CSL), which predict periodic interfaces for particular misorientation. 
As such, they are defined by five macroscopic DoF (directions and rotation), among which some lead to particularly stable configurations with low interfacial energy. Additionally, it is known that considering the microscopic DoF (translation) accessible to any grain boundaries is crucial to fully determine the most stable configurations. However, the NTB we observed does not have so many DoF. In particular, the NTP being confined within a well defined bulk \MAX{} phase, it does not allows the NTB for translation DoF. Allowing the NTB for microscopic DoF might thus lead to lower energy configurations. This is out of the scope of the present work, which is focused on the characterization of the NTP we observed experimentally.

For large angles $\Psi$, the NTB cannot be conveniently described by dislocations anymore as discussed above and illustrated in the insets of Fig.~\ref{fig:InterfaceEnergy}. 
We propose that for high-angles of rotation, the NTB can be appropriately represented by disclination loops. As the rotational counterpart of dislocations, disclinations are line defects which introduce a discontinuity of the elastic rotation field, referred to as the Frank vector. The concept of disclinations is appropriate for the description of elastic fields of tilt boundaries with high misorientations~\cite{FRESSENGEAS2014} and for nanotwin microstructures~\cite{REINHOLZ2016,CAPOLUNGO2019}. 
Regarding the cookie shape NTP considered in this work, the stress field due to the network of overlapping dislocation arrays can be equivalently and more simply described by using two disclination loops delimiting the NTP. Such representation was succesfully applied for nanolamellar twins with tilt misorientations~\cite{CAPOLUNGO2019}. Here, we propose to apply this concept to twist interfaces with high misorientations, which has never been attempted so far to our knowledge. More details can be found as Supplementary Material.

As shown in Fig.~\ref{fig:DisclinLoops}(a), a NTP with misorientation $30^{\circ}$ is modelled by a dipole of twist disclination loops with Frank vector magnitude $30^{\circ}$ about the Z (\hkl[0001]) axis. A recent field disclination mechanics framework is used to built this loop dipole~\cite{FRESSENGEAS2014,CAPOLUNGO2019}. The two main Cauchy stress tensor components predicted by the model are shear stresses $\sigma_{xz}$ and $\sigma_{yz}$. The latter are characteristic of screw dislocation out of plane shear stresses and clearly arise from the twist rotation discontinuity applied within the NTB. 

Fig.~\ref{fig:DisclinLoops}(b) shows $\sigma_{xz}$ and $\sigma_{yz}$ as predicted by atomistic simulations and evidenced a fair match with what is predicted by the disclination model in Fig.~\ref{fig:DisclinLoops}(a). 
The similarities are 
(1)~the morphology of the surrounding stress field maxima with two inverse dipoles in a particular crystallographic direction, and 
(2)~the magnitude of these stress dipoles around \SI{\pm 1}{\giga \pascal}. 
This match between atomistic and continuum representation is however partial. The differences are 
(1)~the direction of stress maxima rotated by \ang{90}, and 
(2)~stresses within the NTB predicted by MD but not by the disclinations.
These differences clearly originate from characteristics not considered by the disclination based model. Typically, dislocation cores effects are only considered by atomistic modelling and leads to $\epsilon_{zz}$ eigenstrains.
Additionally, the reconstruction of the NTB for high twist angles we observed by MD, appears to play a crucial role in the distribution of the stress field. Disclination based models will have to be enriched to consider such atomistic mechanisms, but they are yet much promising to model general interfaces in complex materials at the continuum scale.

Regarding the formation mechanism of the NTP, our preliminary observations suggest a strong influence of the Ti-Al diffusion in the basal planes at the onset of the NTB. The diffusion could be favoured by the interfacial dislocation network that locally induce a variation of free volume within basal planes.

The exact impact of a nano-twist phase on the properties of the \MAX{} nanolayered structure is yet unclear. But it seems that they are not anecdotal events, as reported in \cite{Drouelle2020Microstructure-oxidationPhase,Zhang2016}.
Similarly to what has been observed with the formation of Frank partial dislocations~\cite{Yu2021FrankDiffusion}, the interfacial dislocation network could limit the propagation of \hkl<a> dislocations and thus participate to the strengthening of \MAX{} phases.
The prismatic interfaces of the NTP would require more investigations as they should play an important role by hindering the propagation of basal dislocation more efficiently than the basal interfaces.
From a more general perspective, nano-twist crystals have been shown to exhibits peculiar optical properties \cite{Lee2021TunableAngle}.

Additional investigations are ongoing to gain a comprehensive understanding on the formation mechanisms of this nano-twist phase, in order to opens up new possibilities for tailored properties MAX phases.

\section*{Acknowledgments}

This project has received financial supports from the CNRS through the MITI interdisciplinary programs and from the National Natural Science Foundation of China (no. 52175284). This  work  was  performed  using HPC resources from GENCI-TGCC (grant 2020-A0080911390 and 2021-A0100911390) and from the EXPLOR center of the Université de Lorraine.

\bibliographystyle{elsarticle-num} 

\bibliography{references_antoine, references_vincent, references_julien}%

\end{document}


\begin{frontmatter}

\title{Features of a nano-twist phase in the nanolayered \MAX
\\
--
\\
Supplementary Material}

\author[lem3,labex]{Julien Guénolé}
\author[lem3,labex]{Vincent Taupin}

\affiliation[lem3]{organization={Université de Lorraine, CNRS, Arts et Métiers, LEM3},
            city={Metz},
            postcode={57070},
            country={France}}

\affiliation[labex]{organization={Labex DAMAS, Université de Lorraine},
            city={Metz},
            postcode={57070},
            country={France}}

\author[lms,ls]{Maxime Vallet}
\affiliation[lms]{organization={Laboratoire Mécanique des Sols, Structures et Matériaux, CentraleSupélec, CNRS UMR 8579, Université Paris-Saclay},
            city={Gif-sur-Yvette},
            postcode={91190},
            country={France}}
\affiliation[ls]{organization={Laboratoire Structures, Propriétés et Modélisation des Solides, CentraleSupélec, CNRS UMR 8580, Université Paris-Saclay},
            city={Gif-sur-Yvette},
            postcode={91190},
            country={France}}         

\author[cmse]{Wenbo Yu}
\affiliation[cmse]{organization={Center of Materials Science and Engineering, School of Mechanical and Electronic Control Engineering, Beijing Jiaotong University},
            city={Beijing},
            postcode={100044},
            country={China}}  

\author[lem3,labex]{Antoine Guitton\corref{cor1}}
 \cortext[cor1]{antoine.guitton@univ-lorraine.fr}

\end{frontmatter}

\section{Experimental sample}

\begin{figure}[hb]
\includegraphics[width=0.99\columnwidth]{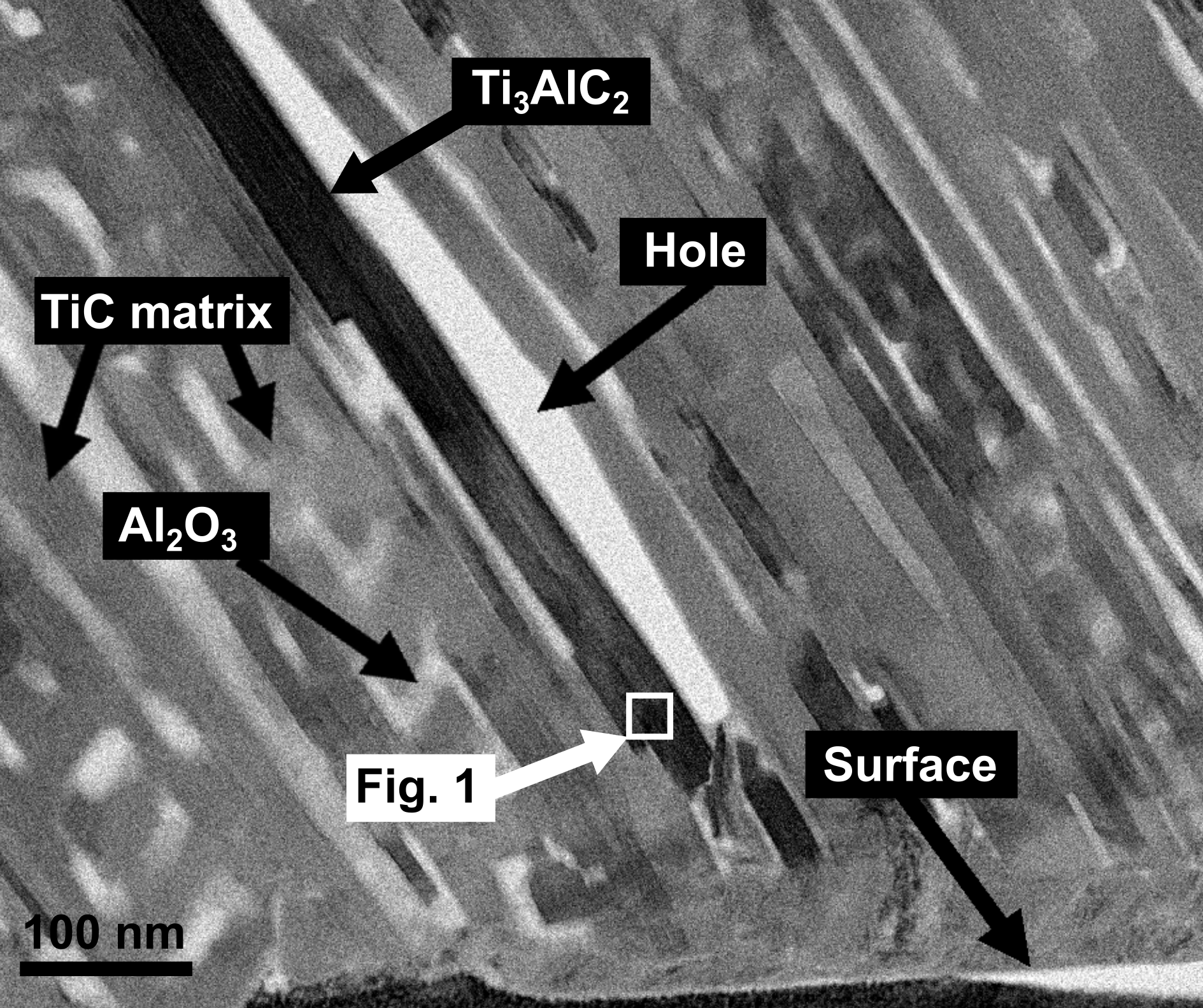}
\caption{\label{fig_SM_TEM} Low magnification micrography of the experimental sample shown in Figure~1 (white empty square). Bright-field STEM micrography of Ti$_3$AlC$_2$ embedded in TiC matrix. Note the presence of Al$_2$O oxide and of a large void at the vicinity of the Ti$_3$AlC$_2$ grain. }
\end{figure}

Fig.~\ref{fig_SM_TEM} shows a low magnification micrography of the experimental sample presented in the manuscript as Figure~1. This bright-field STEM micrography evidences an extended Ti$_3$AlC$_2$ grain embedded in TiC matrix. The presence of Al$_2$O oxide and of a large void at the vicinity of the Ti$_3$AlC$_2$ grain is also visible. The area indicated by a white empty square corresponds to the filtered HR-STEM micrography in HAADF mode, namely Figure~1, in the main manuscript.

\section{Disclination model of the nano-twist phase}

\noindent  The main aspects of the disclination model is presented here. For more details, the reader is referred to papers mentioned in the main text and references therein. A small deformation framework is chosen. The material displacement vector $\bf u$ is a single-valued continuous field. The
distortion tensor is the gradient of the displacement
${\bf U} = {\bf grad \hspace{1mm} u}$. It is a curl-free field:

\begin{equation} \label{curlU}
{\bf curl \hspace{1mm} U} = 0.
\end{equation}

\noindent The strain field $\boldsymbol \epsilon$ is the symmetric part of $\bf U$, while the
rotation field $\boldsymbol \omega$ is the skew-symmetric part. The rotation vector $\vec{\boldsymbol \omega}$ reads:

\begin{equation}\label{xomega}
\vec{{\boldsymbol \omega}} = \frac{1}{2} {\bf curl\, u}.
\end{equation}

\noindent This rotation field is also a single-valued and continuous quantity. The gradient of the rotation vector is the curvature tensor ${\boldsymbol \kappa}$:

\begin{equation}\label{curv}
{\boldsymbol \kappa} = {\bf grad} \vec{{\boldsymbol \omega}}. 
\end{equation}

\noindent It is also called the bend-twist tensor. Like the material distortion ${\bf U}$, the material curvature tensor is a curl-free tensor. To introduce disclinations in the body, we use the deWit's disclination density tensor ${\boldsymbol \theta}$. It has components $\theta_{ij}=\Omega_i t_j$ in the cartesian frame $({\bf e}_x, {\bf e}_y, {\bf e}_z)$. The components correspond to a Frank vector angle $\Omega_i$ about ${\bf e}_i$ per unit resolution surface $\Delta S$, for a line vector component along direction ${\bf e}_j$, $t_j$. Disclinations are indeed line defects, like dislocations, but they introduce a discontinuity of the elastic (or plastic, up to a sign) rotation that corresponds to the Frank vector. Considering a finite surface $S$ of unit normal 
${\bf n}$, which is thread by an ensemble of disclination lines corresponding to a non-zero disclination density, the Frank vector reads:

\begin{equation}\label{Frankvector1}
{\boldsymbol \Omega} =  \int_S {\boldsymbol \theta} \cdot \bf{n} {\it dS}.
\end{equation}

\noindent Because of the elastic/plastic rotation discontinuity the plastic ${\boldsymbol \kappa}_p$ and elastic ${\boldsymbol \kappa}_e$ parts building the total material curvature contain incompatible
components that are not curl-free. The incompatibility of elastic and plastic curvatures writes:

\begin{equation}\label{incompk}
{\bf curl} \hspace{1mm} {\boldsymbol \kappa}_e = - {\bf curl } \hspace{1mm}
{\boldsymbol \kappa}_p = {\boldsymbol \theta}.
\end{equation}

\noindent Although it is not the case in the present simulations, dislocations can also be introduced in addition to disclinations. Such is done by using the Nye's dislocation density tensor ${\boldsymbol \alpha}$. It has components $\alpha_{ij}=B_i t_j$, corresponding to a Burgers vector length $B_i$ along direction ${\bf e}_i$ per unit resolution surface $\Delta S$, and to a dislocation line vector component $t_j$ along direction ${\bf e}_j$. Due to the elastic/plastic displacement discontinuity, ${\it i.e.}$ the Burgers vector, the plastic ${\boldsymbol \epsilon}_p$ and elastic ${\boldsymbol \epsilon}_e$ strains building the total material strain contain incompatible, non curl-free components, such that:

\begin{eqnarray} \label{alphamod}
{\bf curl} \hspace {1mm} {\boldsymbol \epsilon}_e &=& +
{\boldsymbol \alpha} + {\boldsymbol \kappa}_e^t - tr({\boldsymbol
\kappa}_e){\bf I} \label{alphamod1} \\
{\bf curl} \hspace {1mm} {\boldsymbol \epsilon}_p &=& -
{\boldsymbol \alpha} + {\boldsymbol \kappa}_p^t - tr({\boldsymbol
\kappa}_p){\bf I}. \label{alphamod2}
\end{eqnarray}

\noindent The above incompatibility equations show how elastic strains are related to the presence of dislocations and disclination-induced elastic curvatures. The elastic strains and curvatures must satisfy the higher-order quasi-static balance equation:

\begin{equation} \label{Mindlin}
{\bf div} \hspace{1mm} {\boldsymbol \sigma}+ \frac{1}{2} {\bf curl}
\hspace{1mm} {\bf div} \hspace{1mm} {\bf M}= 0.
\end{equation}

In the above equation, ${\boldsymbol \sigma}$ is the Cauchy stress tensor. The tensor ${\bf M}$ is called couple stress tensor and is the work-conjugate of curvatures, just like the stress tensor is that of strains. Usually, the elastic constitutive relations are taken in the form:

\begin{eqnarray}
{\boldsymbol \sigma} &=& {\bf C} : {\boldsymbol \epsilon}_e + {\bf D} :
{\boldsymbol
\kappa}_e  \label{T} \\
{\bf M} &=& {\bf A} : {\boldsymbol \kappa}_e + {\bf B} :
{\boldsymbol \epsilon}_e. \label{M}
\end{eqnarray}

\noindent In the above ${\bf A}$, ${\bf B}$, ${\bf C}$ and ${\bf D}$ are elasticity tensors. In the present simulations, only the classical tensor ${\bf C}$ is considered such that couple stresses are neglected. This choice does not alter the predicted internal stresses due to disclinations. The elasticity moduli are in Voigt notation in the hexagonal crystal frame $C_{11} = C_{22} = \SI{353}{\giga\pascal}$, $C_{12} = \SI{75}{\giga\pascal}$, $C_{13} = \SI{69}{\giga\pascal}$, $C_{33} = \SI{296}{\giga\pascal}$ and $C_{44} = C_{55} = C_{66} = \SI{119}{\giga\pascal}$. All equations are numerically solved with a spectral code taking advantage of Fast Fourier Transform (FFT) algorithms. The incompatibility equations are solved in the Fourier space with the use of centered difference schemes for the evaluation of spatial derivatives. The stress equilibrium equation is solved by using the accelerated scheme together with the use of the rotated scheme for the evaluation of the modified Green operator. In the simulation shown in the main text, the FFT grid size is $256 \times 256 \times 32$ voxels in the X, Y and Z directions. The voxel size is $\SI{0.4}{\nano\meter}$ in all directions. Periodic boundary conditions are used in all directions, like in molecular dynamics simulations. \\

\begin{figure}[t]
\includegraphics[width=0.9\columnwidth]{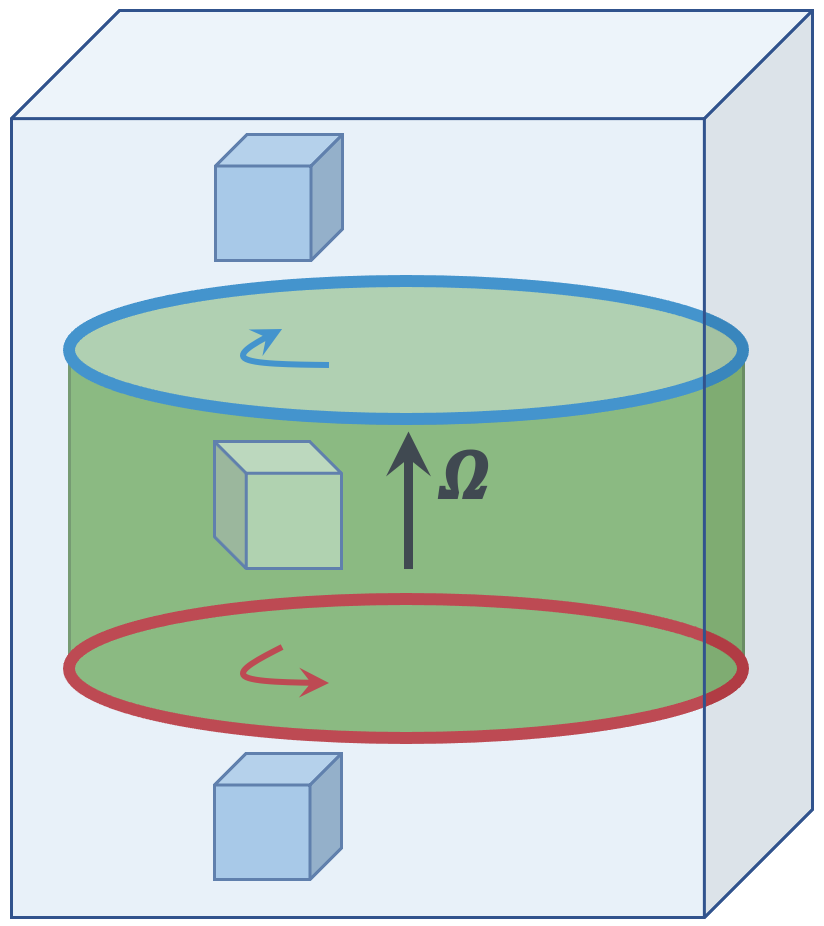}
\caption{\label{DisclinLoops} Illustration of the use of the disclination model to build a nano-twist phase. See the text for details.}
\end{figure}

The figure (\ref{DisclinLoops}) explains how the disclination concept is used to build a nano-twist, penny-shaped, phase. The big cube is the matrix material. In this matrix the crystal is unrotated, as sketched by the two small cube orientations near the bottom and the top surfaces. To start building the nano-twist phase, we first introduce a positive twist disclination loop near the bottom surface. The Frank vector is shown in the figure, it is perpendicular to the surface delimited by the disclination loop. By crossing upwards the surface delimited by the disclination loop, the elastic rotation encounters a jump equal to the Frank vector of the disclination loop. The jump of crystal rotation is shown by the rotating arrow inside the disclination loop, and the new orientation of the crystal is shown by the small cube in the middle of the material. The crystal is twisted with respect to the matrix phase. Finally, to finish building the nano-twist phase, we must add a negative disclination loop near the top surface, with the same Frank vector magnitude, such that when crossing the surface delimited by this disclination loop, the crystal rotation encounters an opposite jump and recovers the initial value in the matrix. The two disclination loops are built with appropriate disclination density components ${\theta}_{zx}$ and ${\theta}_{zy}$ spread on a few voxels and such that the Frank vector magnitude according to equation (\ref{Frankvector1}) is 30 degrees. Note that elastic stresses and strains are continuous and not singular close to disclination lines because we use continuous disclination densities.  \\